\documentstyle[12pt,epsfig]{article}
\def\np{\vspace{12pt} \noindent}

\begin{document}

\begin{center}
{\LARGE \bf {Detection of the velocity dipole in the radio galaxies of the NRAO VLA Sky Survey}}
\end{center}
\begin{center}
{\large \bf {Chris Blake \& Jasper Wall}}
\end{center}
\begin{center}
{\large \bf {Astrophysics, Nuclear and Astrophysics Laboratory, \\ Keble Road, Oxford, OX1 3RH, UK}} \\
{\bf {E-mail: cab@astro.ox.ac.uk}}
\end{center}

{\np \bf
We are in motion against the cosmic backdrop.  This motion is evidenced by the systematic temperature shift -- or dipole anisotropy -- observed in the Cosmic Microwave Background radiation (CMB) \cite{smo}.  Because of the Doppler effect, the temperature of the CMB is 0.1 per cent higher in our direction of motion through the Universe.  If our standard cosmological understanding is correct, this dipole should also be present as an enhancement in the surface density of distant galaxies \cite{ellbal}.  The main obstacle in finding this signal is the very uneven distribution of nearby galaxies in the Local Supercluster, which drowns out the small cosmological imprint.  Here we report the first detection of the expected dipole anisotropy in the galaxy distribution, in a survey of galaxies detected in radio waves.  Radio galaxies are mostly located at cosmological distances, so the contamination from nearby clusters should be small.  With local radio sources removed, we find a dipole anisotropy in the radio galaxy distribution in the same direction as the CMB, close to the expected amplitude.  This result is confirmation of the standard cosmological interpretation of the CMB.
}

\np
The dipole anisotropy in the galaxy distribution due to our motion -- the ``velocity dipole'' -- has two causes.  Firstly, emissions from galaxies ahead of us are Doppler boosted with respect to those behind, enhancing their fluxes; hence more are detected above a given flux threshold.  Secondly, aberration of angles (as predicted by special relativity) causes the position vectors of galaxies to be displaced towards our direction of motion.  We must distinguish the velocity dipole -- an imprint in the distribution of distant, evenly distributed galaxies -- from the anisotropic distribution of local galaxies due to the Local Supercluster.  We describe the latter as the ``clustering dipole''; it will not be a dipole distribution, but will have a dipole component.  The clustering dipole is the cause of our motion through the Universe (the gravitational pull of the Local Supercluster), whilst the velocity dipole is the consequence of it.  To measure the velocity dipole we must remove the contamination of the clustering dipole.

\np
The velocity dipole is a small effect, causing a surface density enhancement of $\sim 1$ per cent.  In order to detect it we must identify a sufficient quantity of distant objects and maintain uniformity in calibration to better than 1 per cent over the sky.  Active galactic nuclei (AGN) are situated in galaxies at cosmological distances and are therefore ideal probes of the velocity dipole.  Unsuccessful attempts have been made to detect it in AGN X-ray surveys \cite{sch} and radio surveys \cite{bal}.  In contrast, the clustering dipole has been detected and quantified in studies probing the local distribution of galaxies at both optical \cite{lah} and infra-red \cite{row} wavelengths.  In particular, Rowan-Robinson et al. \cite{row} used the Infra-Red Astronomical Satellite (IRAS) PSCz catalogue to compute the implied motion of the Local Group of galaxies.

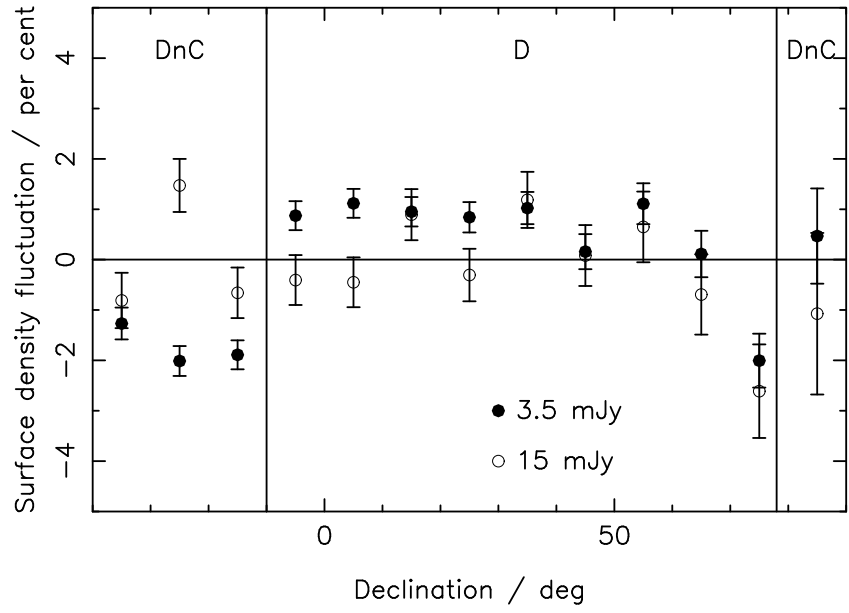
\begin{figure}[h]
\begin{center}
\caption{}
\epsfig{file=surf.ps,width=8cm,angle=-90}
\label{figsurf}
\end{center}
\end{figure}

\np
We measure the galaxy dipole using the NRAO VLA Sky Survey of radio sources (NVSS) \cite{con}.  The NVSS covers a large fraction of the celestial sphere (82 per cent) and is deep ($\sim 50$ sources deg$^{-2}$), thus providing sufficient sources for good statistics.  Moreover, the radio AGN in the survey are at a median redshift $\overline{z} \sim 1$ \cite{dunpea}, so that the contamination from local sources should be minimal.  To prevent pollution from Galactic radio sources, we restrict our analysis to Galactic latitudes greater than $15^o$.  The expected dipole amplitude is so small that systematic fluctuations above 1 per cent over wide angles will drown dipole signal.  At flux-densities below 15 mJy, the NVSS does suffer from spurious systematic fluctuations in source surface density of this magnitude (Figure \ref{figsurf}).

\np
To eliminate the clustering dipole we remove from the sample radio sources within 30 arc-seconds of known nearby galaxies as listed in the IRAS PSCz catalogue \cite{sau} and the Third Reference Catalogue of Bright Galaxies (RCBG3) \cite{cor}.  The IRAS galaxies eliminate local spiral galaxies obeying the far-infrared/radio correlation, and the RCBG3 sources remove radio sources powered by AGN in local elliptical galaxies.

\np
To measure the dipole from the remaining distribution of galaxies, we measure the spherical harmonics of the distribution.  Any density field across the sky $\sigma(\theta,\phi)$ -- where $\theta$ ($0 \rightarrow \pi$) and $\phi$ ($0 \rightarrow 2\pi$) are ordinary spherical polar co-ordinates -- can be expanded as a sum of spherical harmonics $Y_{l,m}(\theta,\phi)$ with coefficients $A_{l,m}$:
\begin{equation}
\sigma(\theta,\phi) = \sum_{l=0}^{\infty} \sum_{m=-l}^{+l} A_{l,m} \, Y_{l,m}(\theta,\phi)
\label{sigdef}
\end{equation}
The $l$th spherical harmonics produce fluctuations on angular scales $\sim 180^o/l$ and are described by $2l+1$ independent coefficients.  A dipole distribution across the sky will produce signal in the three $l=1$ spherical harmonic coefficients, but no signal in the others.  To estimate the coefficients $A_{l,m}$ from $N$ discrete objects with positions $(\theta_i,\phi_i)$, we calculate
\begin{equation}
A_{l,m} = \sum_{i=1}^N Y_{l,m}^* (\theta_i,\phi_i)
\label{clmdef}
\end{equation}

\np
We measure the harmonic coefficients $A_{l,m}$ of the NVSS galaxy distribution for $1 \le l \le 3$ using equation \ref{clmdef} (a total of 15 independent coefficients).  We measure the $l>1$ coefficients because, as the sky is incomplete (containing blank regions below declination $-40^o$ and inside Galactic latitude $15^o$), higher harmonics are influenced by a dipole distribution.  Further, any calibration gradients would (in general) bias all harmonics.  Having measured the coefficients, we then determine how well they are fit by a dipole model (with the same blank regions).

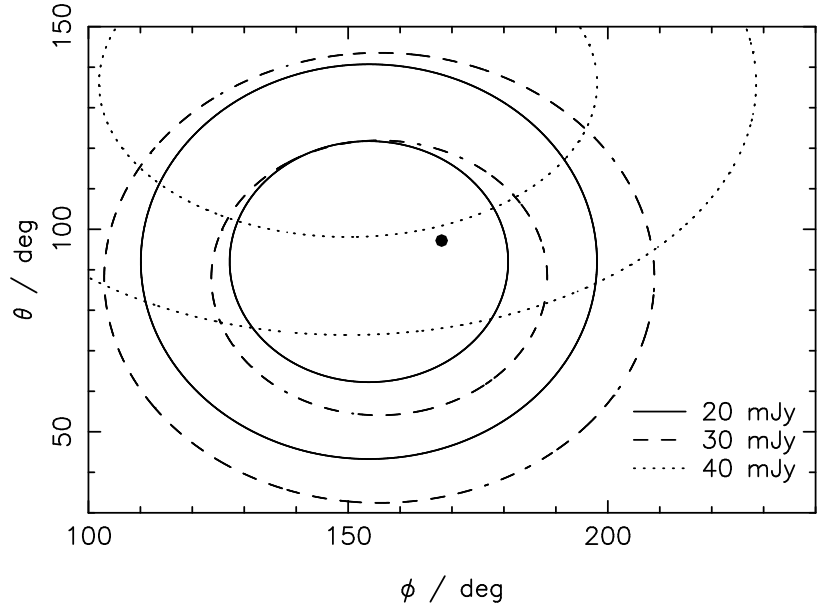
\begin{figure}[h]
\begin{center}
\caption{}
\epsfig{file=dir.ps,width=8cm,angle=-90}
\label{figdir}
\end{center}
\end{figure}

\np
A dipole model has three parameters: an amplitude $\delta$ and direction described by two angles $(\theta,\phi)$.  We define $\delta$ as the total fractional fluctuation in surface density from maximum to minimum.  Table \ref{tabres} displays the best-fitting dipole parameters and errors for a sequence of flux-density thresholds and describes the method used.  The dipole model is a good fit ($\chi_{red}^2 \sim 1$) down to 15 mJy.  Below this level the NVSS data-reduction biases of Figure \ref{figsurf} become significant, producing a sudden increase in $\chi_{red}^2$ and a significant shift in the best-fitting $\theta$ (but not $\phi$), anticipated because the biases are declination-dependent.  Figure \ref{figdir} illustrates the best-fitting direction $(\theta,\phi)$ of the measured dipole at intensity levels above 20 mJy.  The direction does not change with flux-density threshold and is in good agreement with the direction of the CMB velocity dipole $\theta = 97.2 \pm 0.1^o, \phi = 168.0 \pm 0.1^o$ \cite{lin}.  The local clustering dipole typically contributes $\Delta \delta = 0.5 \times 10^{-2}$ to the total amplitude.

\np
Ellis \& Baldwin \cite{ellbal} derived the expected amplitude of the velocity dipole assuming radio sources to have identical power-law spectra $S(\nu) \propto \nu^{-\alpha}$ and an integral source-count $N(>S) \propto S^{-x}$:
\begin{equation}
\delta_{pred} = 2 \, (v/c) \, [ \, 2 + x(1+\alpha) \, ]
\end{equation}
where $v$ is the peculiar velocity of our reference frame with respect to the frame in which the radio source population is (assumed) isotropic, and $c$ is the speed of light.  CMB dipole measurements imply $v = 370 \pm 2$ km s$^{-1}$ \cite{lin}.  Taking $x \approx 1$ and a mean spectral index $\alpha \approx 0.75$, we find $\delta_{pred} = 0.9 \times 10^{-2}$.  Note that $v$ is the peculiar velocity of the Sun in the CMB frame.  (The speed of the Earth around the Sun is $\sim 30$ km s$^{-1}$ and can be neglected).  Our observed values of $\delta$ are on average $1.5 \, \sigma$ away from this prediction.

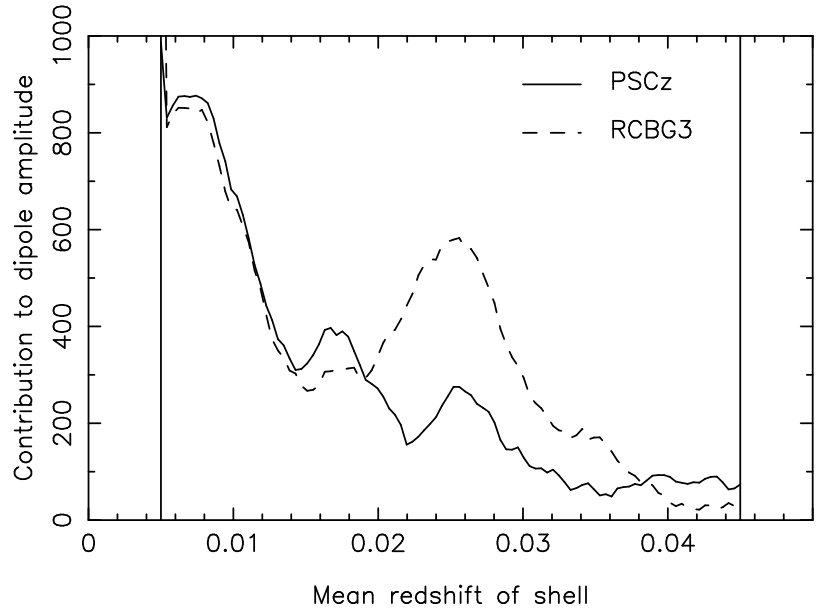
\begin{figure}[h]
\begin{center}
\caption{}
\epsfig{file=dipred.ps,width=8cm,angle=-90}
\label{figdipred}
\end{center}
\end{figure}

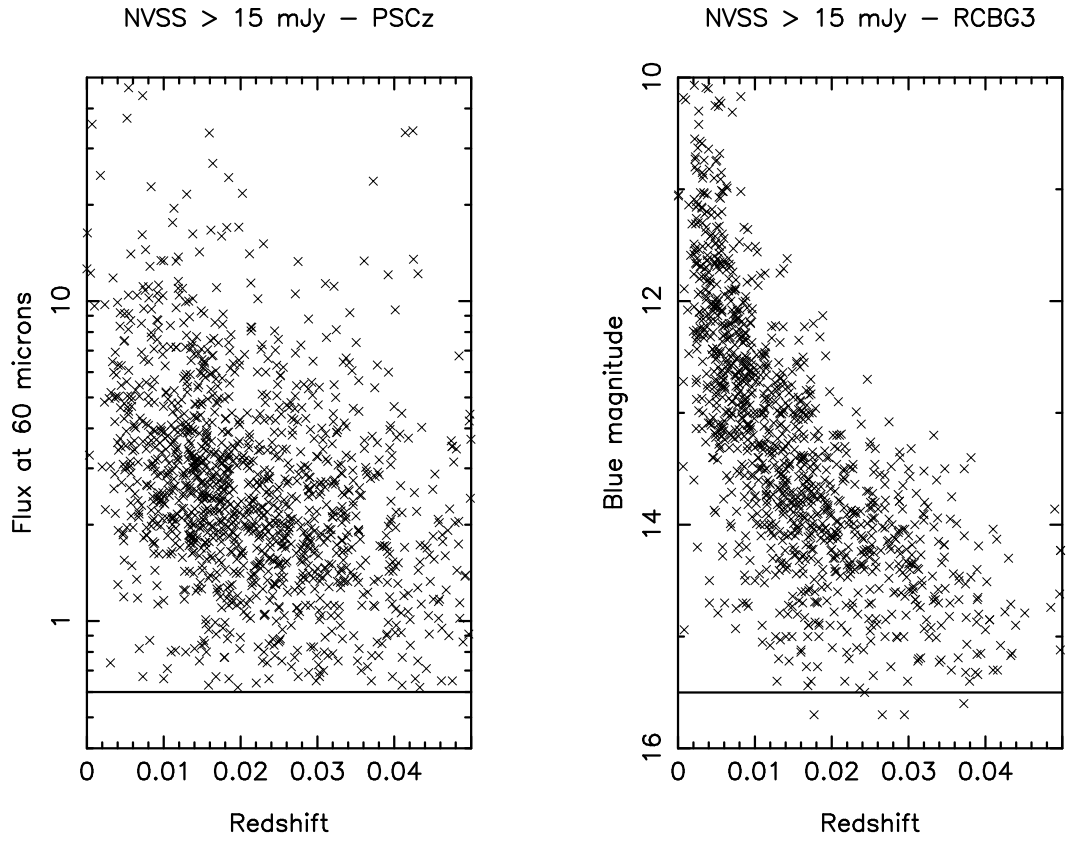
\begin{figure}[h]
\begin{center}
\caption{}
\epsfig{file=magred.ps,width=11cm,angle=-90}
\label{figmagred}
\end{center}
\end{figure}

\np
We can be confident that our measured surface-density dipole is due to velocity rather than residual local clustering.  Almost all the contribution to the local clustering dipole comes from redshifts $z < 0.03$ -- analysis of the IRAS PSCz dipole \cite{row} shows that the dipole amplitude has almost converged by $100 \, h^{-1}$ Mpc.  This result is verified in Figure \ref{figdipred}, where we plot the strength of the dipole contribution in redshift shells from the PSCz and RCBG3 catalogues.  Furthermore, these catalogues contain almost all the radio galaxies to $z = 0.03$.  This is demonstrated by Figure \ref{figmagred} which gives magnitude-redshift plots for PSCz and RCBG3 galaxies matched with NVSS sources.  The distributions make it clear that we are missing a negligible fraction of radio galaxies at $z < 0.03$.

\np
It is hard to conceive of any other effect that can produce our measured dipole.  A signal originating from instrumental or data reduction effects would tend to skew the dipole direction towards the north or south poles, as such effects only depend on declination.  However, our measured dipole points to declination $\approx 0^o$ in agreement with the CMB dipole.  Moreover instrumental effects would (in general) produce signal in higher (non-dipole) harmonics, but the best-fit dipole model is a good fit with $\chi_{red}^2 \sim 1$.  Furthermore, the dipole amplitude is extremely unlikely to be a chance fluctuation of an isotropic universe.  For example, for a flux-density threshold of 20 mJy, the isotropic universe ($\delta = 0$) model fits the observed harmonics with $\chi_{red}^2 = 2.28$, rejecting this model at a confidence level of $> 99.5$ per cent.  It is difficult to see what else could produce the dipole, particularly one pointing (within the $1\sigma$ direction contours) in the CMB dipole direction.  For the 20 mJy measurement, the $1\sigma$ confidence region for the direction encompasses just 5 per cent of the sky.

\np
The alignment of our measured velocity dipole with the CMB dipole supports the predictions of fundamental assumptions about the cosmological origin of the CMB, the lynchpin of much present-day observational cosmology.  This result indicates that objects at $z \sim 1$ (such as distant radio galaxies) can define the cosmic frame and have an isotropic distribution in that frame.

\np
{\small{We are grateful to Lance Miller and Steve Rawlings for comments on early drafts.  We also acknowledge the referee Jim Condon for detailed and helpful criticism.}}

\pagebreak

\begin{table}
\begin{center}
\caption{}
\label{tabres}
\vspace{24pt}
\begin{tabular}{cccccc}
\hline
Flux & $N$ & Best $\delta$ & Best $\phi$ & Best $\theta$ & $\chi_{red}^2$ \\
/ mJy & & $\times 10^{-2}$ & / $^o$ & / $^o$ & \\
\hline
$> 40$ & 125,603 & 1.4 $\pm$ 0.9 & 149 $\pm$ 49 & 135 $\pm$ 38 & 1.02 \\
$> 35$ & 143,524 & 1.7 $\pm$ 0.7 & 161 $\pm$ 44 & 117 $\pm$ 39 & 0.74 \\
$> 30$ & 166,694 & 2.2 $\pm$ 0.7 & 156 $\pm$ 32 & 88 $\pm$ 33 & 1.01 \\
$> 25$ & 197,998 & 2.2 $\pm$ 0.6 & 158 $\pm$ 30 & 94 $\pm$ 34 & 1.01 \\
$> 20$ & 242,710 & 2.1 $\pm$ 0.6 & 153 $\pm$ 27 & 93 $\pm$ 29 & 1.32 \\
$> 15$ & 311,037 & 1.5 $\pm$ 0.5 & 148 $\pm$ 29 & 59 $\pm$ 31 & 1.81 \\
$> 10$ & 431,990 & 1.0 $\pm$ 0.4 & 132 $\pm$ 29 & 25 $\pm$ 19 & 4.96 \\
\hline
\end{tabular}
\end{center}
\end{table}

\np
{\bf Table \ref{tabres}: The dipole amplitude and direction that best fit the galaxy distribution.}  The Table lists the best-fitting dipole model parameters $(\delta,\theta,\phi)$ together with $1\sigma$ errors for NVSS sources above various flux thresholds, after local sources have been removed.  $\delta$ is expressed as a percentage.  $N$ is the number of sources left in the unmasked survey region and the reduced $\chi^2$ values of the fit to the observed survey harmonics are also shown.  The expected (CMB) velocity dipole parameters are $\delta = 0.9 \times 10^{-2}$, $\phi = 168^o$, $\theta = 97^o$.  To find the best-fitting parameters, we vary the values of $(\delta,\theta,\phi)$ over a grid.  For each set of parameters, we randomly generate a dipole galaxy distribution and mask it in the same way as the real data (below declination $-40^o$ and inside Galactic latitude $15^o$).  We then measure the $A_{l,m}$'s of this test distribution, averaging over 100 realizations to increase accuracy.  We hence obtain a mean and standard deviation for each $A_{l,m}$ in the set corresponding to the parameters $(\delta,\theta,\phi)$.  We can then compute the $\chi^2$ statistic with the observed survey harmonics.  The best-fitting dipole parameters are found by minimizing $\chi^2$; this minimum value (goodness-of-fit) indicates how well the dipole model fits the data.  The errors on the best-fitting parameters are properly described by $\chi^2$ contours in the parameter space.  For example, with all three parameters varying, the $1\sigma$ and $2\sigma$ error regions are enclosed by $\chi^2$ increases of 3.53 and 8.02.

\pagebreak

\np
{\bf Figure \ref{figsurf}: The NVSS suffers from systematic fluctuations in source surface density across the sky at low flux-densities.}  Here we plot the surface density as a function of declination, in declination bins of width $10^o$, for flux-density thresholds 3.5 mJy (filled circles) and 15 mJy (open circles).  The declination range of each array configuration (D or DnC) is shown.  The 2 per cent fluctuations present at the completeness limit 3.5 mJy largely disappear when the threshold is raised to 15 mJy.  These biases result from the sparse {\it uv}-plane sampling of the NVSS (W. Cotton, private communication, 2001).  The fluctuations are a function of declination only, as the NVSS beam and analysis procedure have no way of knowing the right ascension of objects.  The NVSS was carried out with the Very Large Array at 1.4 GHz in D and DnC configurations during 1993 to 1996.  The survey covers all the sky north of declination $-40^o$; the catalogue contains $\sim 1.8 \times 10^6$ sources and is claimed to be $\sim 99$ per cent complete at integrated flux density 3.5 mJy.  The Full-Width Half-Maximum of the synthesized beam is $\sim 45$ arcsec.  Prior to analysis, we masked 22 regions around bright and extended radio galaxies; these can appear in the NVSS catalogue as clusters of many point sources by virtue of the source-finding algorithm.  These regions constitute a tiny fraction of the surveyed area and were filled with random distributions of sources at the mean surface density of the rest of the survey.  The large NVSS beam means that only a small fraction of galaxies are resolved into double sources -- this is important because a large number of doubles would increase the source surface density and hence the observed dipole amplitude.

\np
{\bf Figure \ref{figdir}: Our measured dipole direction agrees with that of the CMB dipole.}  Here we plot $1\sigma$ and $2\sigma$ contours for the best-fitting dipole direction for flux-density thresholds of 40 mJy (dotted), 30 mJy (dashed) and 20 mJy (solid).  These contours are defined by $\Delta\chi^2 =  2.30$ and $6.17$ (with $\theta$ and $\phi$ varying and $\delta$ fixed at its best-fitting value).  The filled circle illustrates the direction of the CMB dipole, which is consistent with our measurements.

\np
{\bf Figure \ref{figdipred}: The clustering dipole originates from redshifts $z < 0.03$.}  Here we measure the strength of the clustering dipole in the PSCz and RCBG3 catalogues as a function of redshift.  At each redshift $z$, we average over a redshift range $\Delta z = 0.01$ and thus the plot starts at $z = 0.005$.  For each redshift shell, the $y$-axis plots $\sqrt{D^2 - N}$ where $D = | \sum_i \hat{r}_i |$ for the galaxies in that shell \cite{bal}.  This expression is proportional to the imbalance in the total number of sources across the sky, the quantity to which the total dipole amplitude is sensitive. The peak in RCBG3 at $z = 0.025$ is due to the Coma cluster.  Hence it seems that once we get past Coma, the local clustering dipole is negligible.

\np
{\bf Figure \ref{figmagred}: The PSCz and RCBG3 catalogues include almost all nearby NVSS galaxies.}  Here we present magnitude-redshift scatter plots for PSCz galaxies (left-hand panel) and RCBG3 galaxies (right-hand panel) matched with NVSS sources brighter than 15 mJy.  The horizontal lines are the stated completeness limits for the two catalogues.  Clearly the bulk of the radio galaxies lie safely above the completeness limits up to at least $z=0.03$.  Although the number of RCBG3 matches starts to die away at $z \approx 0.025$, this is likely due to the $N(z)$ of the radio population rather than any serious incompleteness in RCBG3 -- doing the same plot for all RCBG3 galaxies shows no such fall-off until the magnitude threshold is reached.

\pagebreak

\pagebreak

\pagebreak

\pagebreak

\begin{thebibliography}{}
\bibitem{smo} Smoot G., Gorenstein M., Muller R., Detection of anisotropy in the cosmic blackbody radiation, Phys. Rev. Lett., 39, 898-901 (1977)
\bibitem{ellbal} Ellis G., Baldwin J., On the expected anisotropy of radio source counts, Mon. Not. R. Astron. Soc., 206, 377-381 (1984)
\bibitem{sch} Scharf C., Jahoda K., Treyer M., Lahav O., Boldt E., Piran T., The 2-10 keV X-ray background dipole and its cosmological implications, Astrophys. J., 544, 49-62 (2000)
\bibitem{bal} Baleisis A., Lahav O., Loan A.J., Wall J.V., Searching for large-scale structure in deep radio surveys, Mon. Not. R. Astron. Soc., 297, 545-558 (1998)
\bibitem{lah} Lahav O., Optical dipole anisotropy, Mon. Not. R. Astron. Soc., 225, 213-220 (1987)
\bibitem{row} Rowan-Robinson M. et al., The IRAS PSCz dipole, Mon. Not. R. Astron. Soc., 314, 375-397 (2000)
\bibitem{con} Condon J.J., Cotton W.D., Greisen E.W., Yin Q.F., Perley R.A., Taylor G.B., Broderick J.J., The NRAO VLA Sky Survey, Astron. J., 115, 1693-1716 (1998)
\bibitem{dunpea} Dunlop J.S., Peacock J.A., The redshift cut-off in the luminosity function of radio galaxies and quasars, Mon. Not. R. Astron. Soc., 247, 19-42 (1990)
\bibitem{sau} Saunders W. et al., The PSCz catalogue, Mon. Not. R. Astron. Soc., 317, 55-64 (2000)
\bibitem{cor} Corwin H., Buta R., de Vaucouleurs G., Corrections and additions to the third reference catalogue of bright galaxies, Astron. J., 108, 2128-2144 (1994)
\bibitem{lin} Lineweaver C., The CMB dipole : the most recent measurement and some history, in Proceedings of the XVIth Moriond Astrophysics Meeting, Gif-sur-Yvette publishers, p.69-75 (1997)
\end{thebibliography}
\end{document}